\documentclass[final,1p,times,authoryear]{elsarticle}
\usepackage{amssymb}
\usepackage{color}
\usepackage{amsmath}
\usepackage{float}        
\usepackage{booktabs}    

\bibpunct{[}{]}{,}{n}{,}{,}

\journal{Nuclear Physics A}

\begin{document}

\begin{frontmatter}

\title{Lifetimes and Transition Probabilities in $N = 76, ^{130}$Xe}

\author[1,2]{D. Kumar}
\ead{devesh.k@vecc.gov.in}
\affiliation[1]{organization={Variable Energy Cyclotron Centre},city={Kolkata}, postcode={700064}, country={India}}
\affiliation[2]{organization={Homi Bhabha National Institute}, city={Mumbai},
postcode={400094},country={India}}
\author[1]{S. Basak}
\author[1,2]{A. Pal}
\author[1,2]{D. Banerjee}
\author[3]{S.S. Alam}
\affiliation[3]{organization={Government General Degree College}, city={Chapra 741123, West Bengal, India}}
\author[4]{S. Rajbanshi}
\affiliation[4]{organization={Presidency University},city={Kolkata}, postcode={700073}, country={India}}
\author[1,2]{T. Bhattacharjee}
\ead{btumpa@vecc.gov.in}

\begin{abstract}
Lifetimes for the ${4}_1^+$ and ${6}_1^+$ states have been measured using $\gamma-\gamma$ fast timing technique for low lying states of $^{130}$Xe, populated from $\beta ^-$ decay of the parent  $^{130}$I produced through $^{nat}U(\alpha,f)$ reaction at $E_{\alpha}$~=~40~MeV. VENTURE array comprising of eight fast CeBr$_3$ detectors is used for the measurement. The reduced transition probabilities deduced from the measured lifetimes have been compared with results from the large basis shell model calculation and the interacting boson model calculation.
\end{abstract}

\begin{keyword}

Nuclear level lifetime \& Transition probabilities \sep $\gamma-\gamma$ fast timing   \sep Shell Model \& IBM 

\end{keyword}

\end{frontmatter}

\section{Introduction}
\label{intro}

Nuclear structure studies around the Z=50, N=82 double shell closure plays a vital role in understanding  nucleon-nucleon interactions and its contribution in the evolution of nuclear shapes ~\cite{Taprogge, vaquero, Jungclaus}. Nuclei lying close to the shell closure are spherical due to strong pairing forces but deformed shapes are preferred as the nucleon numbers deviate from closed shells. This evolution is primarily due to the increase in residual interactions among the valence nucleons~\cite{Jones, Allmond, Gray,  Bally}. The interplay between these two forces are the reason behind observation of shape transition and shape coexistence in the midshell region.

The structure of neutron rich Xe nuclei, located within the proton mid-shell between $Z = 50$ and $Z = 64$, are expected to show the interplay between collective and single particle excitations ~\cite{nomura1,nomura2}. The $R_{4/2}$ ratios and the evolution of transition strengths in the even-even Xe isotopes display the transitional behavior from $\gamma$ soft rotor [O(6) ($R_{4/2} \sim 2.5)$] to spherical vibrator [SU(5) $(R_{4/2} \sim 2.0)$], as one moves towards the neutron shell closure at $N=82$~\cite{Garrel,coquard1,Ilieva,safi}. The electromagnetic properties for the low lying states in lighter even-even $^{124,126}$Xe suggest prolate deformation with a deviation from the O(6) symmetry~\cite{kisyov}.  The systematics of the M1 transition strengths as a function of the mass number hinted for a phase transition near A=130 \cite{Garrel}. The interacting boson fermion model also describes the energy systematics in odd-A Xe isotopes with the combination of spherical and gamma unstable shape and that $^{129-131}$Xe are candidates for shape phase transitions from O(6) to SU(5)~\cite{jafarijadeh}. Initially  $^{128}$Xe was proposed as an example of critical point symmetry (E(5)) for transition between O(6) to SU(5)~\cite{clark}. However, later it was concluded that $^{128}$Xe is not a close realization of the symmetry E(5) and  $^{130}$Xe nucleus was proposed as the most likely candidate for the realization of the E(5) symmetry ~\cite{coquard2}. Recent measurements nullify this latter proposition as well ~\cite{peters}. In fact, quadrupole deformation measurements in the $^{130}$Xe nucleus, performed recently, pointed to the presence of triaxial degree of freedom in the low-lying level structure in $^{130}$Xe ~\cite{Morrison}.

Central to all these observations, understanding the electromagnetic transition strengths and their ratios are required for interpreting the nuclear shape, shape transitions and critical point symmetry features of the quantum phase transition~\cite{Iachello,Casten}. The level structure of $^{130}$Xe is very well known~\cite{Hopke, Mattera}. In $^{130}$Xe, the B(E2) strengths for several transitions are known from Coulomb excitation measurements~\cite{coquard1, Morrison}. The lifetimes for the 2$_1^+$ and  4$_1^+$ levels are known from Doppler-shift data~\cite{Jakob, Konstantinopoulos}. However no direct lifetime measurement data
has been reported for 6$_1^+$ level in this nucleus. Accordingly, direct measurement of lifetimes for the yrast levels in $^{130}$Xe becomes important. 

In the present work, direct measurement of lifetimes have been carried out for the 4$_1^+$ and 6$^+_1$ states in $^{130}$Xe  using $\gamma-\gamma$ fast timing technique~\cite{regis16} with VENTURE array~\cite{venture}. Large basis shell model and interacting boson model (IBM) calculations have been performed to interpret the experimental data.

\section{Experiment}
\label{expt}

\begin{figure}[ht!]
\begin{center}
\includegraphics[width=\columnwidth]{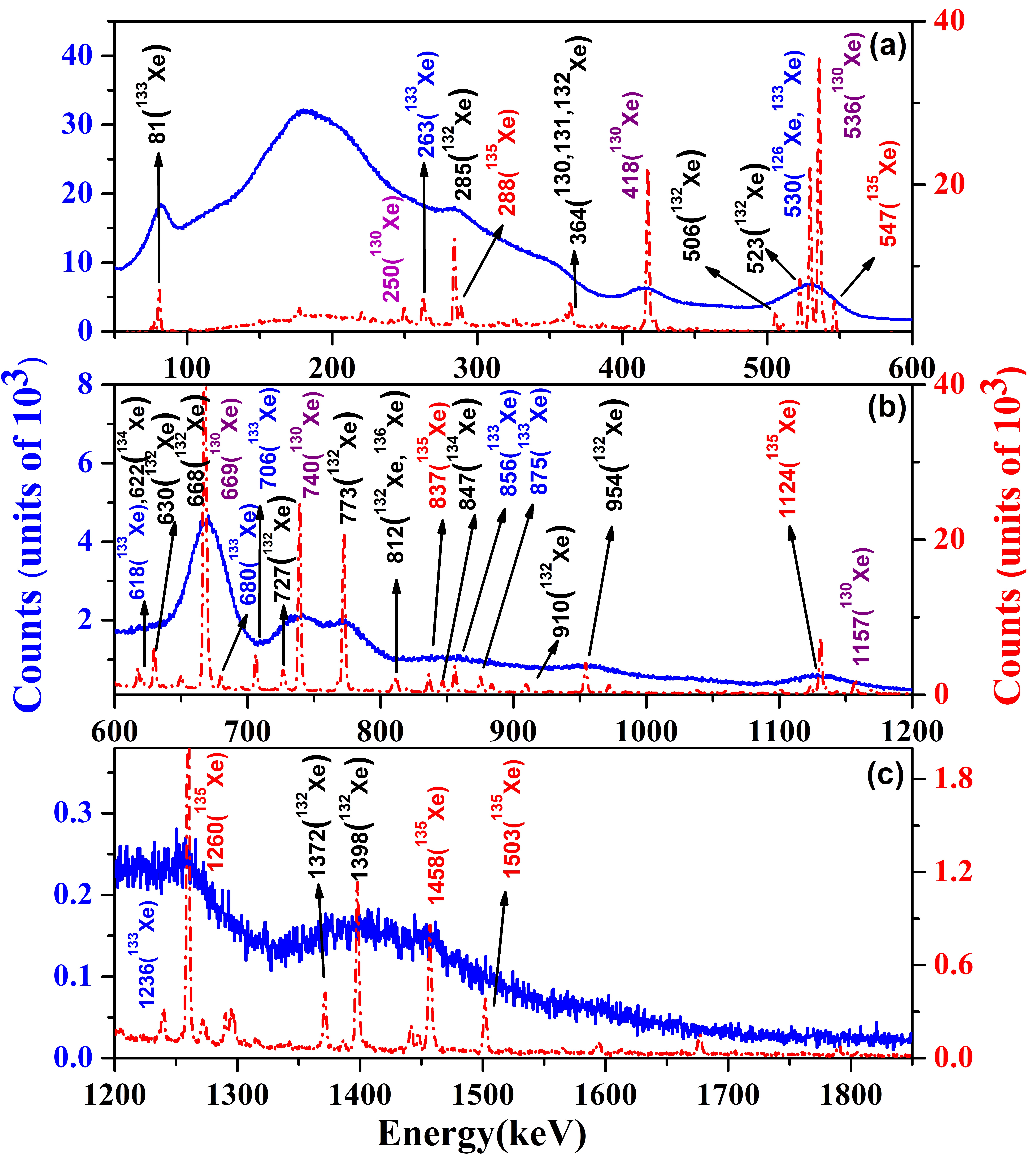}
\caption{ (Color online)  The CeBr$_3$ (blue, solid) and Clover (red, dash-dot-dot) total projections obtained from CeBr$_3$-CeBr$_3$ and CeBr$_3$-Clover coincidence data, respectively. This was used to identify the de-exciting $\gamma$ lines from excited levels of Xe fission products detected in CeBr$_3$ and Clover detectors of the VENTURE setup. The Clover detector spectrum is obtained after Compton suppression which is absent in case of CeBr$_3$ spectrum resulting in large Compton background in the data (blue solid line). The significant difference between the two spectra at low energy is contributed by the interdetector Compton scatterings.}
\label{energy}
\end{center}
\end{figure}

The low-lying excited states of $^{130}$Xe was populated from the $\beta ^-$ decay of neutron rich $^{130}$I~(t$_{1/2}$=12.36 hours) which were produced from $^{nat}U(\alpha,f)$ reaction at E$_{\alpha}$=40~MeV from $K-130$ cyclotron at VECC, Kolkata. The irradiation was done with stacked foil technique~\cite{pm150_cs} with multiple targets inter-spaced with aluminum catcher foils. The targets and catchers were used to radio-chemically separate and extract the iodine isotopes from the bulk of other fission products. The source of separated iodine (liquid with fixed volume of 200 $\mu$l put at the bottom of a vial of dia. 10 mm as a droplet) was subsequently kept at the centre of the VENTURE array~\cite{venture} to count the decay $\gamma$ rays. The array consisted of eight $ 1"~\times~1"$ cylindrical CeBr$_3$ scintillator detectors, coupled to a 2 in. photomultiplier tube (Hamamatsu R9779). In addition,  two Compton (BGO) suppressed Clover HPGe detectors of VENUS array~\cite{npa_arunabha} were also placed for clean identifications of the gamma-rays from fission fragments. Repeated irradiation and measurements were performed with appropriate cooling time for gaining $\gamma-\gamma$ statistics. The CeBr$_3$ total projection from the CeBr$_3$-CeBr$_3$ coincidences is shown in Fig.~\ref{energy} and is compared to the Clover total projection from CeBr$_3$-Clover coincidence events. These total projections are obtained without any subtraction of the underlying background except the Compton suppression obtained in Clover with the use of BGO shields. The comparative spectra could identify the strong $\gamma$ peaks in $^{130}$Xe (parent: $^{130}$I; t$_{1/2}$ = 12.36(1)~h), $^{132}$Xe (parent: $^{132,132m}$I; t$_{1/2}$ = 2.295(13)~h~and~1.387(15)~h, respectively), $^{133}$Xe (parent: $^{133}$I; t$_{1/2}$ = 20.83(8)~h) and $^{135}$Xe (parent: $^{135}$I; t$_{1/2}$ = 6.58(3)~h) without having any contamination from any other fission products.

The Full Width at Half Maximum (FWHM) of the time distribution spectrum corresponding to a prompt $\gamma-\gamma$ cascade ($\tau$< 1~ps), is about 154(8) ps for 1173-1332 keV cascade of $^{60}$Co with a combination of two CeBr$_3$ detectors of the setup whereas it is 188(3) ps for the VENTURE array of 8 detector setup for the same cascade~\cite{venture}. The time-resolution of a $\gamma-\gamma$ can be expressed in term of FWHM by $\sigma_\tau$ = ln(2).FWHM \cite{regis2010}. The Time to Amplitude Converter (TAC) modules were used to measure the time difference distributions between each pairs from the eight detectors of the VENTURE array using {\it `Common Start'} technique~\cite{venture}. The TAC ranges were set to 50~ns and the time distributions were recorded in a 8K histogram so that the time difference can be measured with 6~ps/channel resolution. The energy spectra for each detector were obtained with high resolution spectroscopy amplifiers. The data were gathered using high resolution VME ADCs (Analog to Digital Converter). Details on the characteristics of the VENTURE array, the electronics setup and the $\gamma -\gamma$ fast timing analysis technique used in the present work can be found in Ref.~\cite{venture}.

Lifetime measurement was performed using Generalized Centroid Difference (GCD) method ~\cite{regis16} in which the delayed and anti-delayed time difference distributions for the respective $\gamma-\gamma$ cascades are analyzed with respect to the Prompt Response Difference (PRD) measured for the setup at the same cascade energies. The PRD, representing the prompt time characteristics of the array, is generated from the analysis of $\gamma -\gamma$ fast timing data taken with $^{152}$Eu source. The prompt response of the array depends on the PMT voltages, CFD settings, detector geometry, etc. along with other long-term effects which has been constantly monitored during the experiment. The methodologies for lifetime measurements using GCD technique can be found in our recent works~\cite{safi,shefali,devesh} and is also elaborated in the next section.

\section{Data Analysis and Results}
\label{result}
\begin{figure}[ht!]
\begin{center}
\includegraphics[width=0.8 \columnwidth, height= 12cm]{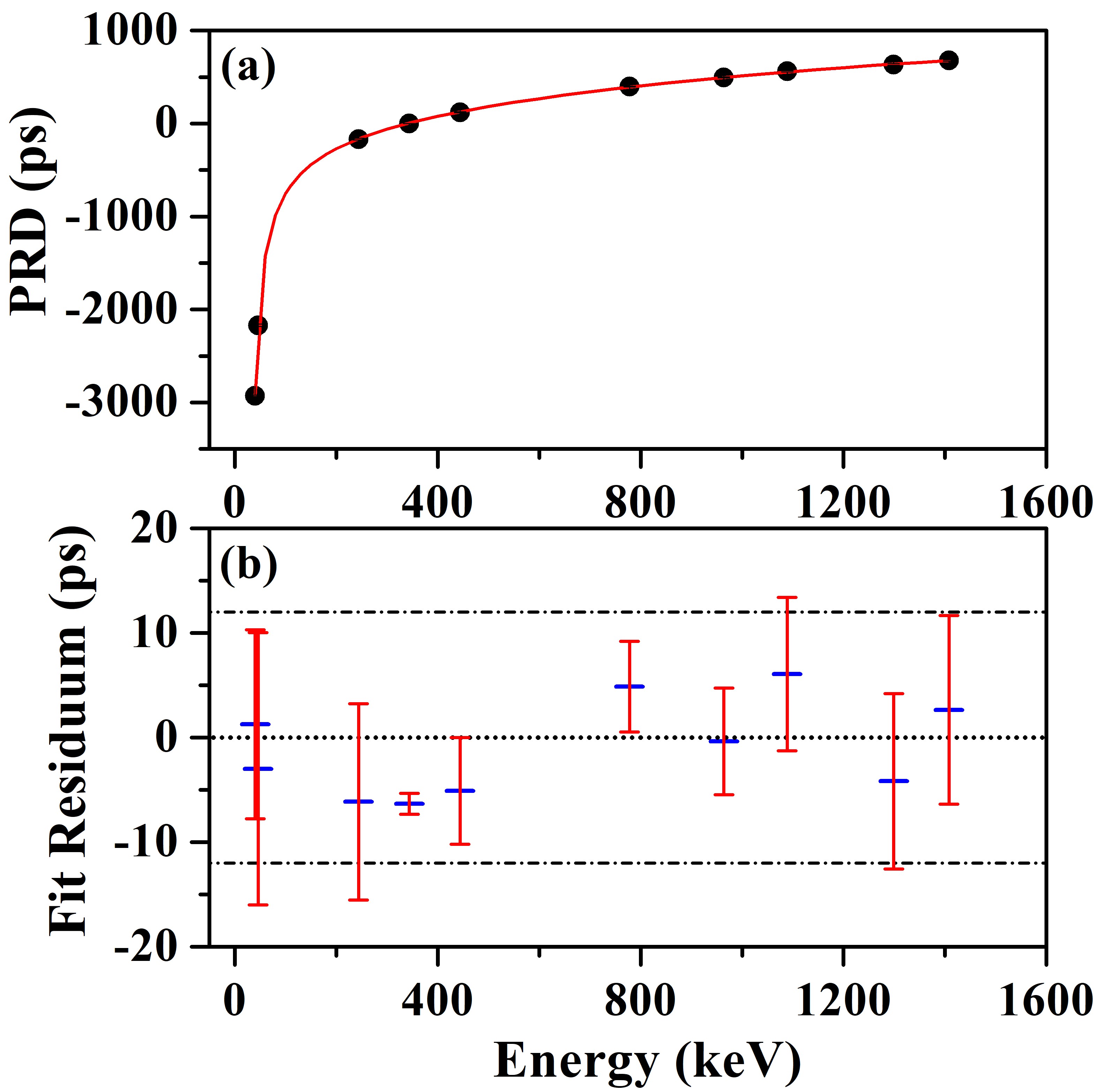}
\caption{{(a)~Prompt Response Difference (PRD) curve obtained for the present experimental setup, showing the time-walk distribution as a function of $\gamma$-ray energy using a $^{152}$Eu source. (b)~Residuals of the PRD fit are shown, with the $3\sigma$ deviation indicated by the dot-dashed lines with respect to the dotted line for zero deviation. A $1\sigma$ value of 4~ps has been considered as the PRD uncertainty ($\delta$PRD) for estimating the error in the measured level lifetime.}}
\label{prd}
\end{center}
\end{figure}

The measurement of nuclear level lifetime ($\tau \pm \delta \tau$) using GCD method can be represented by the following  equation.
\begin{eqnarray}
&&\tau = \frac{1}{2}[\Delta C_{FEP} - PRD] \\ \nonumber \label{eqn1}
\end{eqnarray}

where the $\Delta C_{FEP}$ represents the true centroid differences between the delayed and anti-delayed time difference distributions of a $\gamma-\gamma$ cascade, therefore, considers only  coincidences between two Full Energy Peak (FEP)s. However, the delayed and anti-delayed time difference distributions are experimentally obtained from the analysis of Time to Amplitude Converter (TAC) data that is projected with the selection of $\gamma$ energy gates in the `start' and `stop' detectors of the array. Therefore, as the corresponding FEP selections also include the contributions from the underlying Compton background, the experimental centroid differences ($\Delta C_{exp}$) between the delayed and anti-delayed time difference distributions need to be corrected to obtain $\Delta C_{FEP}$. The said background corrections ($t_{corr}$) are obtained following the methodology described in Ref.~\cite{regis2020}. The following sets of equation explain the determination of $t_{corr}$ and consequently, $\Delta C_{FEP}$ from $\Delta C_{exp}$.

\begin{eqnarray}
\nonumber
&&t_{corr} = \frac{p/b(E_{decay}).t_{corr}(feeder)+p/b(E_{feeder}).t_{corr}(decay)}{p/b(E_{feeder})+p/b(E_{decay})}\\ 
&&{\text \: where}\\ \nonumber
&&t_{corr}(feeder) = [\frac{\Delta C_{exp} - \Delta C_{BG}}{p/b}]_{feeder}\\ \nonumber
&&t_{corr}(decay) = [\frac{\Delta C_{exp} - \Delta C_{BG}}{p/b}]_{decay}\\ 
&&\Delta C_{FEP} = \Delta C_{exp} + t_{corr} \\ \nonumber
\end{eqnarray}

The peak to background ratios $(p/b)$, required to determine the background corrections for feeder and decay ($t_{corr}(feeder),\: t_{corr}(decay)$), are estimated from the gated energy projections corresponding to feeder and decay. The centroid differences, $\Delta$C$_{BG}$, for feeder and decay are determined from the polynomial fitting of the centroid differences obtained from the time difference distributions of corresponding Compton-FEP coincidences. The errors in lifetime $\tau$ are determined from the propagation of errors in different experimentally obtained parameters, viz., $\delta \Delta C_{exp}, \: \delta PRD \: \textrm{and} \: \delta t_{corr}$ using following equation:
\begin{eqnarray}
&&\delta \tau = \frac {1}{2} \sqrt{(\delta \Delta C_{exp})^2 + (\delta t_{corr})^2 + (\delta PRD)^2} \\ \nonumber
\label{eqn2}
\end{eqnarray}

\begin{figure}[ht!]
\begin{center}
\includegraphics[width=0.9\columnwidth, height=12cm]{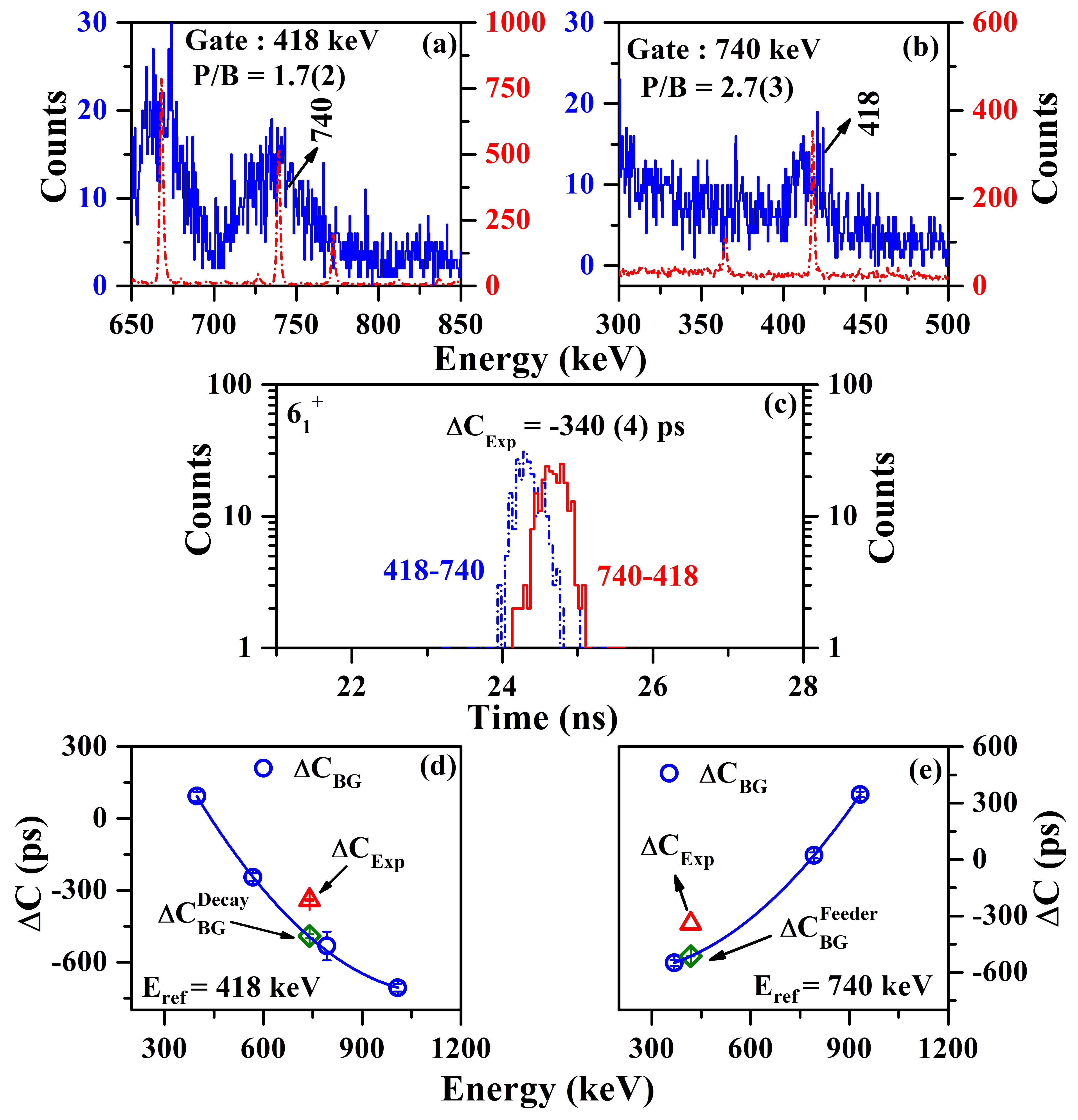}
\caption{The GCD analysis for the ${6}_1^+$ level of $^{130}$Xe is shown. (a) and (b) show the energy gates corresponding to the 418-740 keV cascade from CeBr$_3$-CeBr$_3$ (solid blue) and CeBr$_3$-Clover (dash-dotted red) coincidences; (c) shows the delayed and anti-delayed TACs for the measurement of the centroid difference. (d) and (e) show the background corrections obtained for decay and feeder, respectively.}
\label{61}
\end{center}
\end{figure}
\begin{figure}[ht!]
\begin{center}
\includegraphics[width=0.9\columnwidth, height=12cm]{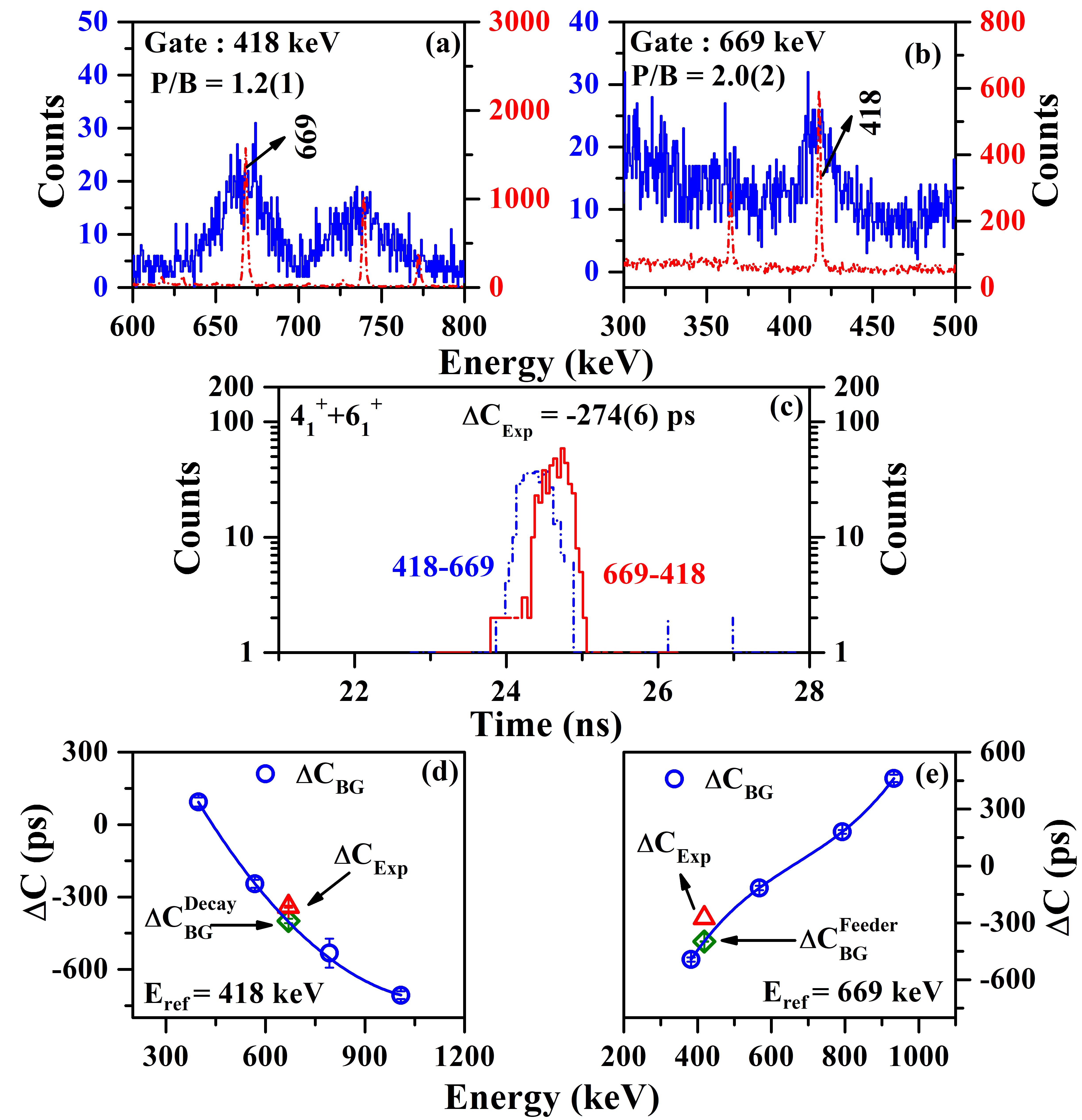}
\caption{The GCD analysis for the ${4}_1^+$+${6}_1^+$level of $^{130}$Xe is shown. (a) and (b) show the energy gates corresponding to the 418-669 keV cascade from CeBr$_3$-CeBr$_3$ (solid blue) and CeBr$_3$-Clover (dash-dotted red) coincidences; (c) shows the delayed and anti-delayed TACs for the measurement of the centroid difference. (d) and (e) show the background corrections obtained for decay and feeder, respectively.}
\label{4161}
\end{center}
\end{figure}
Fig.~\ref{prd} represents the PRD obtained with the experimental seup of VENTURE in the present experiment.  The PRD at low energy is higher due to much higher background compared to earlier works~\cite{venture} arising from significant inter-detector scattering in the present geometry of the setup. However, the energies of interest in the present work are higher than 344~keV where the PRD values are much smaller and well calibrated with many numbers of data points.The PRD uncertainty $\delta PRD$ = 4 ps, corresponding to the standard deviation of the experimental data to the fitted PRD curve, was considered for calculating the error in the measured lifetime following equation~(4).

\begin{table*}[]
\begin{center}
\caption{Lifetime results are shown in bold for the low-lying levels of $^{130}$Xe for which measurements have been carried out in the present work. The relevant centroid differences required for the measurements of lifetimes are also shown.See text for details.}
\centering
\resizebox{\textwidth}{!}{
\begin{tabular}{ccccccccc}
\hline
\hline
Nucleus& J$^{\pi}$ & cascade & $\Delta C_{exp}$ & t$_{corr}$ & $\Delta C_{FEP}$ & PRD & \multicolumn{2}{c}{Lifetime ($\tau$)} \\
&&&(ps)&(ps)&(ps)&(ps)&\multicolumn{2}{c}{}\\
&&&&(total)&&&{\bf Pres. work} & Lit.~\cite{Jakob, Konstantinopoulos,ensdf}\\
\hline
$^{130}$Xe&6$^+_1$& 418-740 & -340(4) & 84(9) & -256(10) & -270(4) & ${\bf 7(5)~ps}$ & ${< 2}$ ns\\
&4$^+_1$+6$^+_1$ & 418-669 & -274(6) & 88(10) & -186(12) & -220(4) & ${\bf 17(6)~ps}$ & - \\
&4$^+_1$&  &&  &   && ${\bf 10(7)~ps}$ & 3.5(2) ps,3.32(3) ps \\
\hline
\hline
\end{tabular}}
\label{tab1}
\end{center}
\end{table*}

As understood from the total and gated projections, the 418-740 keV $\gamma-\gamma$ cascade is very  clean for lifetime measurement of ${6}_1^+$ state. The delayed and anti-delayed time difference distributions along with the background correction for the lifetime measurement for ${6}_1^+$  state at 1944 keV in $^{130}$Xe are represented in Fig.~\ref{61}.  However, the lifetime of ${4}_1^+$ state couldn't be measured using 740-669 keV cascade due to the contamination from 727-668 keV cascade in $^{132}$Xe. So, the 418-669 keV cascade has been used to measure the added lifetime of ${4}_1^+$ and ${6}_1^+$ levels using "absolute shift measurement"~\cite{devesh, Mach}. The delayed and anti-delayed time difference distributions along with the background correction for the lifetime measurements for ${4}_1^+$+${6}_1^+$ in $^{130}$Xe are represented in Fig.~\ref{4161}. The present measurement yielded the lifetimes for ${4}_1^+$ and ${6}_1^+$ levels to be 10~(7)~ps and 7~(5)~ps, respectively, which are shown in Table~\ref{tab1} with all relevant details and in comparison with the literature values.

\section{Discussion}
\label{disc}

\noindent
The reduced transition probabilities, $B(E2)$, corresponding to the decay of low-lying levels in $^{130}$Xe, have been deduced from the measured lifetimes. For this purpose, the internal conversion coefficients for the respective E2  transitions were calculated using the BrIcc code~\cite{bricc}. The asymmetric uncertainties in the $B(E2)$ values were evaluated by propagating the large experimental uncertainties in the measured lifetimes following Ref.~\cite{barlow}.

\noindent
The observed level structure and electromagnetic transition rates have been interpreted within the framework of large-basis shell-model (LBSM) and interacting boson model (IBM) calculations. 

\noindent
The shell-model calculations were performed using the NuShellX code~\cite{Brown}. For the LBSM calculations, the proton and neutron particles were distributed over the 50–82 major shell comprising the $1g_{7/2}$, $2d_{5/2}$, $2d_{3/2}$, $3s_{1/2}$, and $1h_{11/2}$ orbitals above the $^{100}$Sn core, without truncation. The {\it sn100pn} interaction~\cite{Brown2} was employed to compute the reduced transition probabilities using effective charges ($e_\pi=1.68$, $e_\nu=0.84$), as adopted from Ref.~\cite{Morrison}.

The Interacting Boson Model (IBM-1)~\cite{Iachello2,casten1} describes low-energy collective excitations in even--even nuclei in terms of correlated pairs of valence nucleons treated as $s$ ($L=0$) and $d$ ($L=2$) bosons. The total boson number $N$ is conserved, leading to an underlying $U(6)$ group structure. In IBM-1, proton and neutron bosons, as well as particle and hole degrees of freedom, are treated equivalently. The model admits three limiting dynamical symmetries: $U(5)$ (spherical vibrator), $SU(3)$ (axially deformed rotor), and $O(6)$ ($\gamma$-soft rotor).

The general IBM-1 Hamiltonian is given by
\begin{equation}
H_{\text{IBM}} = \epsilon\, n_d + \kappa\, Q \cdot Q + \kappa'\, L \cdot L + \kappa''\, P \cdot P,
\end{equation}
while the electric quadrupole transition operator is
\begin{equation}
T(E2) = e_b \left[ (d^{\dagger}s + s^{\dagger}\tilde{d}) + \chi (d^{\dagger}\tilde{d})^{(2)} \right],
\end{equation}
where $e_b$ is the effective boson charge and $\chi$ is the structure parameter. Numerical diagonalization of the Hamiltonian was performed using the \textsc{PHINT} code developed by Scholten~\cite{Scholten}, in which the parameters $\epsilon$, $2\kappa$, $2\kappa'$, and $\kappa''$ correspond to the $d$-boson energy (EPS), quadrupole--quadrupole (QQ), angular-momentum (ELL), and pairing (PAIR) interactions, respectively.

For the nucleus $^{130}$Xe ($Z=54$, $N=76$), the experimental excitation-energy ratios $R_{4/2}=E(4_1^+)/E(2_1^+)=2.25$ and $R_{6/2}=E(6_1^+)/E(2_1^+)=3.63$~\cite{Morrison,Hopke,Mattera} lie between the $U(5)$ and $O(6)$ limits. This intermediate character identifies $^{130}$Xe as a strong candidate for exhibiting critical-point $E(5)$ symmetry, as proposed by Iachello~\cite{Iachello2000} and further discussed in Ref.~\cite{DaLiYuXin2003}.

Within the IBM framework, the $E(5)$ symmetry corresponds to the critical point of the second-order shape-phase transition between the $U(5)$ and $O(6)$ dynamical symmetries, as defined using the coherent-state formalism~\cite{Dieperink1980,CastenZamfir2000}. To describe this transitional behavior for a finite boson number, calculations were performed assuming $N_B=5$ and employing the reduced Hamiltonian
\begin{equation}
H_{\text{IBM}} = \epsilon\, n_d + \kappa''\, P \cdot P,
\end{equation}
which is sufficient to track the evolution of $^{130}$Xe along the $U(5)$--$O(6)$ leg of the nuclear shape-phase (Casten) triangle~\cite{Casten,CastenMcCutchan2007}. The parameters were chosen as EPS~=~1056~keV and PAIR~=~36~keV to reproduce both the excitation energies and the $B(E2)$ transition strengths. The calculated transition probabilities were normalized using the weighted mean value of $33.6(7)$~W.u., obtained from the three previously measured experimental $B(E2;2_1^+ \rightarrow 0_1^+)$ values listed in Table~\ref{be2_comparison}.

\begin{figure}[ht!]
\centering
\includegraphics[width=0.6\textwidth, height=8cm]{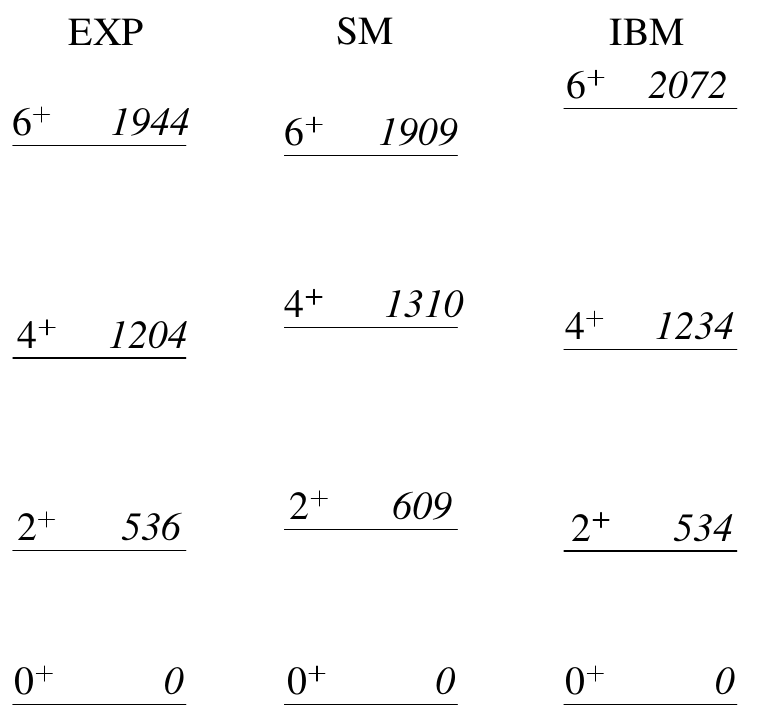}
\caption{ Comparison of the experimentally observed low-lying excited levels of $^{130}$Xe with those calculated using the shell model and IBM approaches.}
\label{SM_IBM}
\end{figure}

The low-lying excitations of $^{130}$Xe and their corresponding calculated energies from the LBSM and IBM calcultions are shown in Fig.~\ref{SM_IBM}. The calculated wave functions reproduce the overall trend of experimental excitation energies, with deviations within $\sim$100~keV, indicating that the adopted shell-model interaction and the IBM parameters capture the essential features of the residual proton–neutron correlations.

 The decomposition of the total angular momentum into proton and neutron components ($I_\pi \otimes I_\nu$) for each level, along with the dominant particle partitions contributing to the configuration-mixed states obtained from shell-model calculation is given in Table~\ref{130Xe_config}. Only those $I_\pi$–$I_\nu$ couplings contributing more than 5\% to the total wave function are listed. For each coupling, partitions with probabilities greater than 1\% are also shown wherever applicable. This information highlights the principal configurations and their relative proton–neutron contributions to the structure of the low-lying states in $^{130}$Xe.

\begin{table}[h!]
\centering
\caption{Experimental and calculated excitation energies and dominant proton–neutron configurations for selected states in $^{130}$Xe. The decomposition of total angular momentum into proton and neutron components ($I_\pi \otimes I_\nu$) is shown, together with the principal partitions contributing to the configuration-mixed shell-model wave functions. Only couplings contributing more than 5\% to the total wave function are listed; for each coupling, partitions with probabilities greater than 1\% are shown where available. Wherever the partitions were completely fragmented, no major partitions are shown.}
\label{130Xe_config}
\renewcommand{\arraystretch}{1.2}
\setlength{\tabcolsep}{5pt}
\resizebox{\columnwidth}{!}{
\begin{tabular}{ccccccccc}
\toprule
\hline
\textbf{Nucleus} & $E_x^{\text{expt}}$ (keV) & $E_x^{\text{calc}}$ (keV) & $J^\pi_{\text{expt}}$ & $J^\pi_{\text{calc}}$ & $I_\pi$ & $I_\nu$ & \% Prob. (a) & Major partition \& \% Prob.  \\
\midrule
$^{130}$Xe & 0 & 0 & $0^+$ & $0^+$ & $0^+$ & $0^+$ & 55 & $\pi(1g_{7/2}^4)\otimes\nu(2d_{3/2}^{-2}1h_{11/2}^{-4})$ (3) \\
 &  &  &  &  &  &  &  & $\pi(1g_{7/2}^4)\otimes\nu(2d_{3/2}^{-2}3s_{1/2}^{-2}1h_{11/2}^{-2})$ (4)  \\
 &  &  &  &  &  &  &  & $\pi(1g_{7/2}^22d_{5/2}^2)\otimes\nu(2d_{3/2}^{-2}1h_{11/2}^{-4})$ (3)  \\
 &  &  &  &  &  &  &  & $\pi(1g_{7/2}^4)\otimes\nu(2d_{3/2}^{-4}1h_{11/2}^{-2})$ (2)  \\
 &  &  &  &  &  &  &  & $\pi(1g_{7/2}^4)\otimes\nu(2d_{5/2}^{-2}2d_{3/2}^{-2}1h_{11/2}^{-2})$ (2)  \\
 &  &  &  &  & $2^+$ & $2^+$ & 38 & $\pi(1g_{7/2}^4)\otimes\nu(2d_{3/2}^{-2}1h_{11/2}^{-4})$ (1) \\
\midrule
 & 536 & 609 & $2^+$ & $2^+$ & $0^+$ & $2^+$ & 37 & $\pi(1g_{7/2}^4)\otimes\nu(2d_{3/2}^{-2}1h_{11/2}^{-4})$ (2) \\
  &  &  &  &  &  &  &  & $\pi(1g_{7/2}^22d_{5/2}^2)\otimes\nu(2d_{5/2}^{-2}2d_{3/2}^{-2}1h_{11/2}^{-2})$ (2)  \\
 &  &  &  &  & $2^+$ & $0^+$ & 27 & $\pi(1g_{7/2}^4)\otimes\nu(2d_{3/2}^{-2}1h_{11/2}^{-4})$ (1) \\
 &  &  &  &  & $2^+$ & $2^+$ & 12 & --- \\
\midrule
 & 1204 & 1310 & $4^+$ & $4^+$ & $0^+$ & $4^+$ & 19 & $\pi(1g_{7/2}^22d_{5/2}^2)\otimes\nu(2d_{3/2}^{-2}1h_{11/2}^{-4})$ (1) \\
 &  &  &  &  & $2^+$ & $2^+$ & 32 & $\pi(1g_{7/2}^4)\otimes\nu(2d_{3/2}^{-2}1h_{11/2}^{-4})$ (1) \\
  &  &  &  &  &  &  &  & $\pi(1g_{7/2}^22d_{5/2}^2)\otimes\nu(2d_{3/2}^{-2}1h_{11/2}^{-4})$ (1)  \\
 &  &  &  &  & $2^+$ & $4^+$ & 6 & --- \\
 &  &  &  &  & $4^+$ & $2^+$ & 8 & --- \\
  &  &  &  &  & $4^+$ & $0^+$ & 19 & --- \\
\midrule
 & 1944 & 1909 & $6^+$ & $6^+$ & $6^+$ & $0^+$ & 34 & --- \\
 &  &  &  &  & $4^+$ & $2^+$ & 21 & --- \\
&  &  &  &  & $2^+$ & $4^+$ & 7 & --- \\
 &  &  &  &  & $0^+$ & $6^+$ & 1 & --- \\
\bottomrule
\hline
\end{tabular}}
\end{table}

\noindent
The configuration analysis reveals that the low-lying states of $^{130}$Xe are primarily governed by neutron-hole excitations within the $h_{11/2}$ and $d_{3/2}$ orbitals, coupled to proton configurations dominated by $(1g_{7/2})^4$ and mixed $(1g_{7/2})^2(2d_{5/2})^2$ components. The collective nature of the low-lying levels is also indicated from the strong configuration mixing observed for these levels. It is also evident that proton excitations are more dominant in generating the higher angular momentum states.
\noindent
A comparison with $^{132}$Xe~\cite{safi} indicates that as neutrons are removed from the $N=82$ closed shell, configuration mixing increases significantly, leading to enhanced collectivity in the low-lying states. 
\begin{table*}[h!]
\centering
\caption{Reduced transition probabilities $B(E2; J_i^+ \rightarrow J_f^+)$ in $^{130}$Xe obtained from the present work, compared with previous experimental data and theoretical predictions. All $B(E2)$ values are in Weisskopf units (W.u.).}
\label{be2_comparison}
\renewcommand{\arraystretch}{1.2}
\setlength{\tabcolsep}{8pt}
\resizebox{\textwidth}{!}{
\begin{tabular}{lccccc}
\toprule
\textbf{Transition} & \multicolumn{5}{c}{$B(E2; J_i^+ \rightarrow J_f^+)$ (W.u.)} \\
\cmidrule(lr){2-6}
 & Present (Exp.) & Previous (Exp.) & LBSM (Present) & IBM (Present) & Theory(Lit.) \\
\midrule
$2_1^+ \rightarrow 0_1^+$ & -- & 32(3)\cite{Morrison}, 37.1(17)\cite{Jakob}, 32(1)\cite{Konstantinopoulos} & 35 & 33 & 41\cite{Teruya}, 39\cite{Gupta}, 33\cite{Shobani}, 24\cite{Prochniak} \\
$4_1^+ \rightarrow 2_1^+$ & $16^{+36}_{-7}$ & 47(4)\cite{Morrison}, 44.5(20)\cite{Jakob}, 46.4(46)\cite{coquard1}, 47(6)\cite{Konstantinopoulos} & 52 & 43 & 59.6\cite{Teruya}, 58\cite{Shobani}, 46\cite{Prochniak} \\
$6_1^+ \rightarrow 4_1^+$ & $14^{+33}_{-6}$ & 60$^{+14}_{-12}$\cite{Morrison}, 69(9)\cite{coquard1} & 29 & 42 & 62.3\cite{Teruya}, 81\cite{Shobani} \\
\bottomrule
\end{tabular}}
\end{table*}

The reduced transition probabilities obtained from the present experiment and the corresponding theoretical calculations are compared with earlier results in Table~\ref{be2_comparison}. The good agreement between theory and experiment within uncertainties supports the reliability of the adopted effective charges ($e_\pi=1.68$, $e_\nu=0.84$)) and the IBM parameters used in this work. These results confirm that $^{130}$Xe lies in a transitional region between vibrational and $\gamma$-soft collective structures, providing a structural link between the near-spherical $^{132}$Xe and the more deformed lighter Xe isotopes.

\section{Summary}
\label{Summary}

\noindent
The $\gamma$–$\gamma$ fast-timing technique has been employed for the direct measurement of level lifetimes for the ${4}_1^+$ and ${6}_1^+$ states in $^{130}$Xe. The corresponding $B(E2)$ values have been deduced from the measured lifetimes. The lifetime for the ${4}_1^+$ level corroborates earlier measurements, while this work provides the first direct determination of the ${6}_1^+$ lifetime. The deduced $B(E2)$ transition probabilities in $^{130}$Xe are in good agreement with both shell-model and IBM calculations, supporting a picture of moderate collectivity and confirming the transitional character of this nucleus.

\section{Acknowledgment}
\label{Acknowledgment}

The effort of the K-130 cyclotron operation group at VECC, Kolkata, is gratefully acknowledged for providing a good-quality alpha beam. The author expresses sincere gratitude to Dr. S. K. Das (Retd.), RCD (BARC), VECC for his valuable suggestions.

\end{document}